\documentstyle[sprocl,epsfig,axodraw,rotate]{article}
\def\be{\begin{equation}}
\def\ee{\end{equation}}
\def\bea{\begin{eqnarray}}
\def\eea{\end{eqnarray}}

\begin{document}

\title{SLEPTON FLAVOR VIOLATION}

\author{F.~DEPPISCH$^1$, H.-U. MARTYN$^2$, H.~P\"AS$^1$, A.~REDELBACH$^1$, 
R.~R\"UCKL$^1$}

\address{$^1$ Institut f\"ur Theoretische Physik und Astrophysik\\
Universit\"at W\"urzburg\\ D-97074 W\"urzburg\\ Germany}

\address{$^2$ I. Physikalisches Institut, RWTH Aachen, D-52074 Aachen, Germany}


\maketitle\abstracts{If the seesaw mechanism is implemented in
supersymmetric theories 
the heavy neutrinos generate corrections to the slepton mass matrices
which enhance the rates of charged
lepton flavor violating processes.
Here we study lepton flavor violating slepton pair 
production and decay 
at a future $e^+e^-$ linear collider in mSUGRA post-LEP benchmark scenarios. 
We present the results of a background simulation and investigate the 
correlations of the collider signals with the corresponding radiative
lepton decays.
}

In the seesaw model  
three right-handed neutrino singlet fields $\nu_R$
are added to the MSSM particle 
content, with a Majorana mass term and Yukawa couplings $Y_\nu$
to the left-handed
lepton doublets \cite{Hisano:1999fj,Casas:2001sr}.
At energies much below the Majorana mass scale $M_R$, 
electroweak symmetry breaking leads to the neutrino mass matrix
\begin{equation}\label{eqn:SeeSawFormula}
M_\nu = m_D^T M^{-1} m_D = Y_\nu^T M^{-1} Y_\nu (v \sin\beta )^2,
\end{equation}
where $m_D=Y_\nu \langle H_2^0 \rangle$ is a $3\times3$ Dirac mass matrix,
\(\langle H_2^0 \rangle = v\sin\beta\) being the appropriate Higgs v.e.v. with 
\(v=174\)~GeV,
\(\tan\beta =\frac{\langle H_2^0\rangle}{\langle H_1^0\rangle}\), and
$M$ is the Majorana mass matrix.
Thus light neutrino masses are naturally generated if the Majorana mass scale
is much larger than the Dirac mass scale.
The light neutrino mass eigenstates $m_i$ are obtained after
diagonalization of 
$M_{\nu}$ by the unitary MNS matrix $U$.

The other neutrino mass eigenstates are dominantly right-handed with 
masses given by $M$. These neutrinos  
are too heavy to be observed directly. However, they
give rise to virtual corrections 
to the slepton mass matrices, which in turn induce 
lepton-flavor violating effects in the charged lepton sector.
In particular, in the mSUGRA scenarios considered here,
the $6\times 6$ mass matrix of the charged sleptons 
is given by
\begin{eqnarray}
 m_{\tilde l}^2=\left(
    \begin{array}{cc}
        m_{\tilde l_L}^2    & (m_{\tilde l_{LR}}^{2})^\dagger \\
        m_{\tilde l_{LR}}^2 & m_{\tilde l_R}^2
\end{array}
      \right)
 \simeq \tilde{m}_{\rm MSSM} + \left(\begin{array}{cc}
\delta m_L^2 & 0\\      
0 & 0
    \end{array}
      \right),
\end{eqnarray}
where $\tilde{m}_{\rm MSSM}$ includes all lepton-flavor conserving
corrections.
In addition to the latter, the renormalization group
evolution generates off-diagonal terms
in $\delta m_{L}^2$ 
which
in leading-log approximation are given by \cite{Hisano:1999fj}
\begin{eqnarray}\label{eq:rnrges}
  \delta m_{L}^2 &=& -\frac{1}{8 \pi^2}(3m_0^2+A_0^2)(Y_\nu^\dagger L Y_\nu), 
\label{left_handed_SSB}
\end{eqnarray}
where
\(L_{ij}=\ln(M_X/M_{i})\delta_{ij}\),
$M_i~(i=1,2,3)$ being the eigenvalues of the Majorana mass matrix $M$,
and $m_0$ and $A_0$
are
the universal scalar mass and trilinear coupling at the GUT scale $M_X$,
respectively.  
In the above, we have chosen a basis in which 
the charged lepton Yukawa couplings and $M$ are diagonal. 

From (\ref{eqn:SeeSawFormula}) one obtains \cite{Casas:2001sr}, 
\begin{eqnarray}\label{eqn:yy}
Y_\nu^{\dagger} L Y_{\nu}=
\frac{M_R}{v^2\sin^2\beta} U \cdot \textrm{diag}\left(m_1, m_2, m_3\right) 
\cdot
U^{\dagger} \ln \frac{M_X}{M_R}.
\label{eq:yukawa}
\end{eqnarray}
assuming that the Majorana masses are degenerate and the MNS matrix $U$ 
is the only source of CP violation in the lepton sector.
Substitution of the neutrino data  
\cite{Deppisch:2004xv} on $m_i$ and  $U$
in (\ref{eq:yukawa}) and evolution  
to the unification scale $M_X$ provides the necessary input in 
(\ref{left_handed_SSB}).

\begin{figure}[!t] 
     \begin{center}
     \epsfig{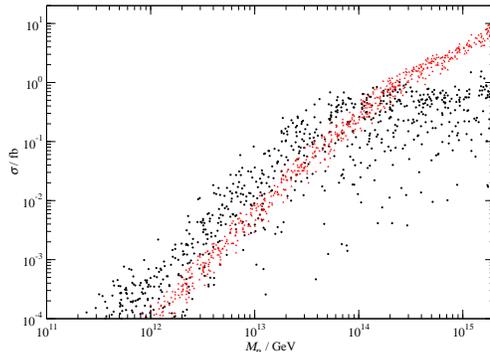}
     \end{center}
     \vspace*{-.5cm}
     \caption{Cross-sections for \(e^+e^- \to \mu^+ e^- +2\tilde\chi_1^0\) 
         (black) and  
         \(e^+e^- \to \tau^+ \mu^- +2\tilde\chi_1^0\) (red) at
         \(\sqrt{s}=500\) GeV as a function of $M_R$ 
         for the mSUGRA scenario SPS1a.
         \label{fig:ep}}
     \end{figure}

\begin{figure}[ht] 
     \begin{center}
     \hspace*{0.5cm}
     \rotate{
     \epsfig{file=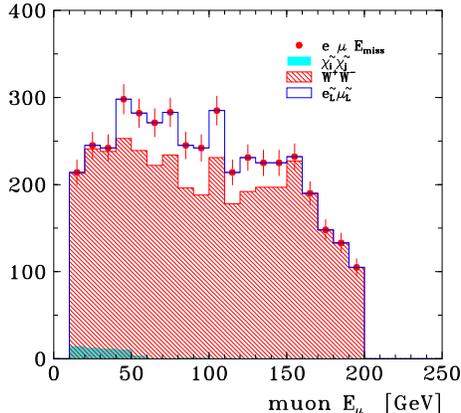,width=5.5cm}}
     \end{center}
     \vspace*{-.7cm}
     \caption{Simulation of $E_\mu$ spectrum
       of $e\mu$ final states in
              scenario SPS1a and SM at $\sqrt{s}=500~{\rm GeV}$, 
              ${\cal L} = 500~{\rm fb}^{-1}$
              and for a LFV signature assuming 
          $\sigma(e^+e^-\to \mu^+ e^- +2\tilde{\chi}^0_1)= 2~{\rm fb}$.      
     \label{fig:emuspec}}
\end{figure}

At low energies, the flavor off-diagonal correction 
(\ref{left_handed_SSB}) induces
the radiative decays 
\(l_i\rightarrow l_j \gamma\).
The decay rates are, schematically, given by \cite{Hisano:1999fj,Casas:2001sr}
\begin{equation}
\Gamma(l_i \rightarrow l_j \gamma) 
\propto \alpha^3 m_{l_i}^5 \frac{|(\delta m_L)^2_{ij}|^2}{\tilde{m}^8} 
\tan^2 \beta,
\label{eq:leg}
\end{equation}
where $\tilde m$ characterizes the sparticle masses in the loop.
At high energies, a feasible test of LFV is provided 
by slepton pair production
$e^+e^- \to \tilde{l}_a^+\tilde{l}^-_b\to 
l_i^+ l^-_j\tilde{\chi}^0_\alpha\tilde{\chi}^0_\beta$, where
LFV can occur in 
production and decay vertices.
For sufficiently narrow slepton 
widths $\Gamma_{\tilde{l}}$ and degenerate slepton masses, the
cross-section can be approximated by
\begin{eqnarray}
\sigma(l_i^+ l_j^-)\propto 
\frac{|(\delta{m}_L)^2_{ij}|^2}{
\tilde{m}^2 \Gamma_{\tilde{l}}^2} \;
\sigma(e^+ e^-\to \tilde{l}^+_a \,
\tilde{l}^-_b) 
Br(\tilde{l}^+_a \to l^+_i \, \tilde{\chi}_0)  
Br(\tilde{l}^-_b \to  l^-_j\, \tilde{\chi}_0),
\label{full_M_squared}
\end{eqnarray}
where \(\sigma(e^+e^-\to \tilde{l}^+_a \,
\tilde{l}^-_b)\) is to be replaced by the 
flavor-diagonal cross-section for slepton pair production. The flavor change
is described by the factor in front of the r.h.s. of (\ref{full_M_squared}).
Our numerical study has been performed for the general case. In particular,
the amplitudes 
for the complete \(2 \to 4\) processes have been summed               
coherently over the intermediate
slepton mass eigenstates. We  
have focussed on 
mSUGRA benchmark scenarios  \cite{Battaglia:2003ab}
designed for linear collider
studies and assumed a very light absolute neutrino mass, $m_1<0.03$~eV.

\begin{figure}[ht] 
     \begin{center}
     \epsfig{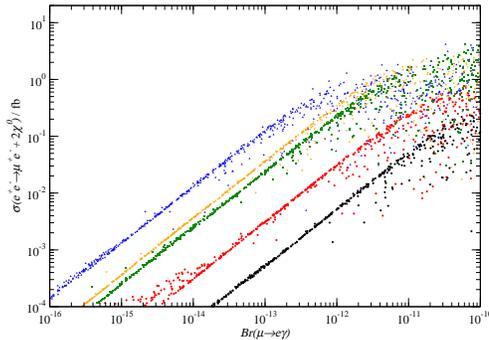}
     \end{center}
     \vspace*{-.5cm}
     \caption{Correlation between \(Br(\mu\to e\gamma)\) and   
     \(\sigma(e^+e^- \to \mu^+ e^- +
     2\tilde\chi_1^0)\) at \(\sqrt{s}=800\) GeV for the mSUGRA 
     scenarios (from left to right) C', G', B', SPS1a  
     and I' {\protect \cite{Battaglia:2003ab}}. \label{fig:emu_lowhigh}} 
\end{figure}

Fig.~\ref{fig:ep} illustrates the dependence 
of the cross-sections\cite{Deppisch:2003wt} for 
\(e^+e^-\rightarrow \mu^+ e^- +2\tilde{\chi}^0_1\) and 
\(e^+e^-\rightarrow \tau^+ \mu^- +2\tilde{\chi}^0_1\)
on the Majorana mass \(M_R\) for the benchmark model SPS1a. 
As can be seen, the cross-sections exhibit the typical proportionality to
$(Y_{\nu} Y_{\nu}^\dagger)$, that is $M_R^2$. The deviation of the behavior
at larger $M_R$ is explained in \cite{Deppisch:2003wt}. 
The scatter plots indicate the impact of 
the uncertainties in the neutrino data.
For a sufficiently large Majorana mass $M_R$ 
the LFV cross-sections can reach several fb. 


The results of a simulation for $\mu e$ final states in the scenario SPS1a
are shown in Fig.~\ref{fig:emuspec}. 
Employing standard selection criteria~\cite{Martyn:2004ew}
the LFV process $e^+e^- \to (\tilde e_L \tilde e_L / \tilde \mu_L \tilde \mu_L)
\to \mu \; e +2 \tilde{\chi}_1^0$
can be detected with $\sim 50 \%$ efficiency. 
The MSSM background can be kept small and is dominated by
chargino and stau production.
The main background comes from SM $W W$ pair production.
Assuming an energy of
$\sqrt{s}=500~{\rm GeV}$ and an integrated luminosity of
${\cal L} = 500~{\rm fb}^{-1}$, a  LFV signal of 1~fb
would result in the observation of a 5 standard deviations effect.
Further improvements are possible, {\it e.g.} by exploiting 
both lepton energy spectra of $\mu$ and $e$
or using polarized beams.
Of particluar interest is the associated $t$-channel production
with right-handed electrons
$e^+e^-_R \to (\tilde e^+_L \tilde e^-_R / \tilde \mu^+_L \tilde e^-_R)
\to \mu^+\tilde{\chi}_1^0 \; e^-\tilde{\chi}_1^0$. 
Although only one slepton contributes to the LFV interaction the signature 
is very clean and the cross section is enhanced, while the $WW$ background is 
suppresed by almost an order of magnitude.
The prospects to identify a LFV 
$\tau \mu$ final state are worse by a factor of about two, because only 
part of the $\tau$ decays can be analysed.

Very interesting and useful are the 
correlations between LFV in radiative decays \cite{Deppisch:2002vz} and 
slepton pair production. 
Such a correlation is illustrated in 
Fig.~\ref{fig:emu_lowhigh} for 
\(e^+e^-\rightarrow \mu^+ e^- +2\tilde{\chi}_1^0\)
and \(Br(\mu\to e \gamma)\).
As expected from the approximations
(\ref{eq:leg}) and (\ref{full_M_squared}),
the neutrino uncertainties drop out, 
while the sensitivity to the mSUGRA parameters remains.
An observation of $\mu \to e \gamma$ with a branching ratio
smaller than the present bound $10^{-11}$ would thus be compatible   
with a cross-section of ${\cal O}(1~{\rm fb})$ for 
$e^+e^- \to \mu^+ e^- +2\tilde{\chi}^0_1$, at least in scenarios 
C', G', B' and SPS1a \cite{Battaglia:2003ab}. 
On the other hand, the non-observation of a signal in the
MEG experiment at PSI would imply the bound $Br(\mu \to e \gamma)<10^{-13}$
and hence predict \(\sigma(e^+e^- \to \mu^+ e^- + 2\tilde\chi_1^0)<
0.1~{\rm fb}\) in the SUSY scenarios considered.

\smallskip
\noindent
This work was supported by the Bundesministerium f\"ur Bildung und 
Forschung (BMBF, Bonn, Germany) under 
the contract number 05HT4WWA2.

\end{document}